%% file: main.tex
\documentclass[10pt,conference]{IEEEtran}

\usepackage[dvipsnames]{xcolor}
\usepackage{algorithmic}
\usepackage{graphicx}
\usepackage{textcomp}
\usepackage{xcolor}

\usepackage{subfiles} 
\usepackage{multirow}
\usepackage{listings}
\usepackage{subcaption}
\usepackage{color,xcolor,colortbl}
\usepackage{xspace}
\usepackage{enumitem}
\usepackage{graphicx}
\usepackage{xr}
\usepackage{url}

\usepackage[most]{tcolorbox} 
\usepackage{booktabs}
\usepackage{diagbox}
\usepackage{amsmath}
\usepackage{array}
\usepackage{colortbl}
\usepackage{color}
\usepackage{balance}
\usepackage{arydshln}
\usepackage{tabularx}

\newcommand\clearrow{\global\let\rowmac\relax}
\clearrow
 \usepackage{adjustbox}
\usepackage[flushleft]{threeparttable}
\usepackage{amsmath, amsfonts}
\usepackage{algorithmic}
\usepackage{graphicx}
\usepackage{textcomp}
\usepackage{xcolor}
\usepackage{xspace}
\usepackage{booktabs}
\usepackage{multirow}
\usepackage{subcaption}
\usepackage{enumitem}
\usepackage{graphicx}  
\usepackage{microtype}
\usepackage{balance}
\usepackage{soul}

\usepackage{hyperref}
\usepackage{hyperxmp}
\usepackage{cleveref}
\usepackage[normalem]{ulem} 

\graphicspath{ {./figures/} }
\AtBeginDocument{%
  \providecommand\BibTeX{{%
    \normalfont B\kern-0.5em{\scshape i\kern-0.25em b}\kern-0.8em\TeX}}}

\def\BibTeX{{\rm B\kern-.05em{\sc i\kern-.025em b}\kern-.08em
    T\kern-.1667em\lower.7ex\hbox{E}\kern-.125emX}}

\definecolor{lightgreen}{rgb}{0.56, 0.93, 0.56}
\definecolor{magicmint}{rgb}{0.67, 0.94, 0.82}
\definecolor{mintgreen}{rgb}{0.6, 1.0, 0.6}
\definecolor{mediumspringgreen}{rgb}{0.0, 0.98, 0.6}
\definecolor{lightgray}{HTML}{EFEFEF}
\definecolor{mycyan}{HTML}{96FFFB}
\definecolor{Lavender}{HTML}{E6E6FA}     
\definecolor{Thistle}{HTML}{D8BFD8}      
\definecolor{Plum}{HTML}{DDA0DD}         
\definecolor{Orchid}{HTML}{DA70D6}       
\definecolor{LightPurple}{HTML}{E6CCFF}  
\definecolor{PaleLilac}{HTML}{EADCF8}   
\definecolor{LightBlue}{HTML}{ADD8E6}     
\definecolor{AliceBlue}{HTML}{F0F8FF}     
\definecolor{PowderBlue}{HTML}{B0E0E6}    
\definecolor{SkyBlue}{HTML}{87CEEB}       
\definecolor{BabyBlue}{HTML}{89CFF0}      
\definecolor{PaleBlue}{HTML}{D6F1FF}      
\definecolor{linecolor}{RGB}{105,105,105} 
\definecolor{bgcolor}{RGB}{234,234,234} 
\definecolor{darkgreen}{rgb}{0.0, 0.5, 0.0}

\usepackage{xspace}
\newcommand{\TOOL}{\text{SE-Jury}\xspace}

\usepackage[normalem]{ulem}

\newcommand*\circled[1]{\tikz[baseline=(char.base)]{
    \node[shape=circle,draw,inner sep=1pt] (char) {#1};}}

\usepackage{array}

\newcommand{\thickcline}[1]{%
  \noalign{\vskip\arrayrulewidth}
  \cline{#1}%
  \noalign{\vskip\arrayrulewidth}
  \cline{#1}%
}

\begin{document}

\title{SE-Jury: An LLM-as-Ensemble-Judge Metric for Narrowing the Gap with Human Evaluation in SE}

\author{
Xin Zhou$^\dagger$,
Kisub Kim$^{\ddagger*}$,
Ting Zhang$^\S$,
Martin Weyssow$^\dagger$,
Lu\'is F. Gomes$^\dagger$, \\
Guang Yang$^\P$,
Kui Liu$^\|$,
Xin Xia$^\#$,
David Lo$^\dagger$ \\
\\
$^\dagger$Singapore Management University, Singapore \\
$^\ddagger$DGIST, Republic of Korea \\
$^\S$Monash University, Australia \\
$^\P$Nanjing University of Aeronautics and Astronautics, China \\
$^\|$Huawei Software Engineering Application Technology Lab, China \\
$^\#$Zhejiang University, China \\
\\
Emails: xinzhou.2020@phdcs.smu.edu.sg, kisub.kim@dgist.ac.kr, ting.zhang@monash.edu, \\
mweyssow@smu.edu.sg, lfgomes@smu.edu.sg, novelyg@outlook.com, brucekuiliu@gmail.com, \\
xin.xia@acm.org, davidlo@smu.edu.sg \\
\\
}

\maketitle

\renewcommand{\thefootnote}{\fnsymbol{footnote}} 
\footnotetext[1]{Corresponding author.}

\begin{abstract}

\subfile{sections/0_abstract}

\end{abstract}

\input{sections/1_introduction}

\input{sections/2_background}

\input{sections/3_approach}

\input{sections/4_setup}

\input{sections/5_results}

\input{sections/6_discussion}
\input{sections/7_related_work}

\input{sections/8_conclusion}

\balance
\bibliographystyle{IEEEtran}
\bibliography{Citation}

\end{document}

%% file: sections/0_abstract.tex
Large Language Models (LLMs) and other automated techniques have been increasingly used to support software developers by generating software artifacts such as code snippets, patches, and comments. However, accurately assessing the correctness of these generated artifacts remains a significant challenge.
On one hand, human evaluation provides high accuracy but is labor-intensive and lacks scalability. 
On the other hand, many automatic evaluation metrics are scalable and require minimal human effort, but they often fail to accurately reflect the actual correctness of generated software artifacts.

In this paper, we present \TOOL, the first evaluation metric for LLM-as-Ensemble-Judge specifically designed to accurately assess the correctness of generated software artifacts. 
\TOOL first defines five distinct evaluation strategies, each implemented as an independent judge. 
A dynamic team selection mechanism then identifies the most appropriate subset of judges as a team to produce a final correctness score through ensembling.
We evaluate \TOOL across a diverse set of software engineering (SE) benchmarks that span three popular SE tasks: code generation, automated program repair, and code summarization. 
Results demonstrate that \TOOL consistently achieves a higher correlation with human judgments, with improvements ranging from 29.6\%–140.8\% over existing automatic metrics. 
\TOOL reaches agreement levels with human annotators that are close to inter-annotator agreement in code generation and program repair. 
These findings underscore SE-Jury's potential as a scalable and reliable alternative to human evaluation in these SE tasks.

%% file: sections/1_introduction.tex
\section{Introduction\label{intro}}

The growing demand to automate software development tasks has led to the emergence of automated techniques for generating software artifacts, such as code snippets~\cite{DBLP:journals/corr/abs-2107-03374,DBLP:journals/corr/abs-2308-10462}, code changes~\cite{DBLP:conf/icse/GuoCXLL0024,DBLP:journals/corr/abs-2303-07221}, and summarization~\cite{DBLP:journals/tosem/SunFCZTYHGHLC24}. 
However, evaluating the correctness of those generated artifacts remains a challenge, largely due to the existence of multiple correct or semantically equivalent solutions for a given problem.

One accurate evaluation method is human evaluation, where multiple human experts directly assess the correctness of the generated artifacts. However, human evaluation is labor-intensive and time-consuming, making it impractical for large-scale assessments. An alternative is test-based metrics, such as pass@k~\cite{DBLP:journals/corr/abs-2107-03374}, where human experts manually design a set of test cases and the generated code is then executed to check whether it passes these test cases. While test-based metrics are more scalable than human evaluation, they still require the careful manual design of comprehensive test cases that cover edge cases~\cite{DBLP:journals/ese/LeTLG18,DBLP:journals/tse/ZhouXKHNLHLL24}. Designing complete test cases is tedious and challenging itself, and many Software Engineering (SE) tasks lack the necessary test cases, making test-based metrics unpractical for large-scale evaluations.

To enable scalable evaluation of generated artifacts with minimal human effort, several automatic evaluation metrics have been proposed. These metrics offer greater scalability by eliminating the need for human evaluation or test cases. However, they are typically less accurate in assessing correctness~\cite{DBLP:journals/jss/EvtikhievBSB23}. 
These automatic metrics can generally be categorized into three types: 1) \textit{Match-based metrics}, 2) \textit{Embedding-based metrics}, and 3) \textit{LLM-as-judge metrics}.
Match-based metrics, such as BLEU~\cite{BLEU} and CodeBLEU~\cite{CodeBLEU}, evaluate the similarity between the generated artifact and a reference, i.e., a correct answer. 
Embedding-based metrics~\cite{BERTScore,CodeBERTScore}, on the other hand, also compare the generated artifact to a reference, but they first encode both into embeddings and then measure the similarity between them. 
In contrast, the LLM-as-judge metric~\cite{ICE-Score} instructs the LLMs to judge the quality of the generated artifact.
Despite the widespread adoption of metrics above, they still suffer from two major \textbf{limitations}.

\vspace{0.1cm}
\noindent\textbf{Interpreting Similarity as Correctness.}
Both match-based and embedding-based metrics use similarity as an indicator of correctness. However, similarity does not always align with correctness. 
For example, if the generated artifact is semantically equivalent to the reference but differs significantly in syntax, the similarity scores would be low, failing to accurately reflect the correctness. 
Moreover, Evtikhiev et al.~\cite{DBLP:journals/jss/EvtikhievBSB23} provided empirical evidence demonstrating a significant misalignment between human judgment and match-based metrics.

\vspace{0.1cm}
\noindent\textbf{Lack of Diverse Evaluation Strategies for Correctness Assessment.}
The representative LLM-as-judge evaluation metric for code, ICE-Score~\cite{ICE-Score}, prompts LLMs to directly assign evaluation scores using instructions such as ``\textit{Assign a score for functional correctness}.'' However, it primarily relies on a single evaluation strategy that involves direct score assignment, without incorporating diverse perspectives to assess correctness. For instance, LLMs can also be prompted to determine whether the generated code is functionally equivalent to a reference implementation or whether it passes a set of test cases. To enable a more reliable and robust evaluation, there is a need for a more comprehensive LLM-as-judge framework that incorporates multiple complementary strategies.

\vspace{0.1cm}
\textbf{Our Work.} 
To address these limitations, we propose \textbf{\TOOL} (\underline{S}oftware \underline{E}ngineering \underline{Jury}), the first LLM-as-Ensemble-Judge metric designed to assess the correctness of generated software artifacts, including code snippets, patches, and comments.
Unlike match-based and embedding-based metrics that approximate correctness through similarity, \TOOL, like other LLM-as-judge approaches, leverages LLMs' reasoning and code comprehension abilities for semantic evaluation.

Inspired by the deliberative process of a legal tribunal composed of domain-specific experts~\cite{inter_court}, where each judge could contribute a unique perspective to reach a well-reasoned verdict, \TOOL employs a multi-evaluator framework. 
Specifically, \TOOL first defines five distinct evaluation strategies, each represented by an independent judge responsible for its own correctness assessment strategy. Second, a team selection mechanism identifies the most suitable subset of judges to form the team. Team selection helps reduce costs by avoiding the use of all strategies for every task and allows the exclusion of strategies that are less suitable for specific tasks.
Third, team members conduct their assessments, and their results are merged into a final correctness score through ensembling.

We evaluate \TOOL on a diverse set of SE datasets, encompassing three popular SE tasks, code generation, automated program repair, and code summarization, across five programming languages: Java, C++, Python, JavaScript, and Go, and cover three different types of generated software artifacts: code snippets, patches, and comments.
Following prior work~\cite{DBLP:journals/jss/EvtikhievBSB23,ICE-Score}, we employ Kendall's $\tau$ coefficient and Spearman's $r_s$ to quantify the statistical correlation between the assessments made by \TOOL and the ground-truth human evaluation results.
The experimental results illustrate that \TOOL achieves substantially and consistently higher correlations 29.6\%–140.8\% than the baselines on average.
Moreover, \TOOL also achieves agreement levels with human annotators that are close to inter-annotator agreement observed in code generation and automated program repair.

\vspace{0.1cm}
\noindent\textbf{Contributions.} The main contributions are as follows:
\begin{itemize}[left=2pt]

    \item To the best of our knowledge, we are the first to advance LLM-as-judge techniques for software by effectively integrating multiple diverse strategies on correctness evaluation.

    \item We propose \TOOL, a novel LLM-as-Ensemble-Judge evaluation framework that enables LLMs to produce correctness assessments that closely align with human judgment. \TOOL achieves this through (1) the introduction of multiple novel evaluation strategies, and (2) a team selection mechanism that effectively combines the strengths of these strategies for robust and reliable evaluation.

    \item Through extensive experiments spanning five programming languages across three popular SE tasks, we demonstrate the effectiveness and generalizability of \TOOL in generating accurate correctness assessments for diverse software artifacts, including code snippets, patches, and comments.
    
\end{itemize}

%% file: sections/2_background.tex
\section{Background\label{background}}

\begin{figure*}[t]
    \centering
    \includegraphics[width=0.75\textwidth]{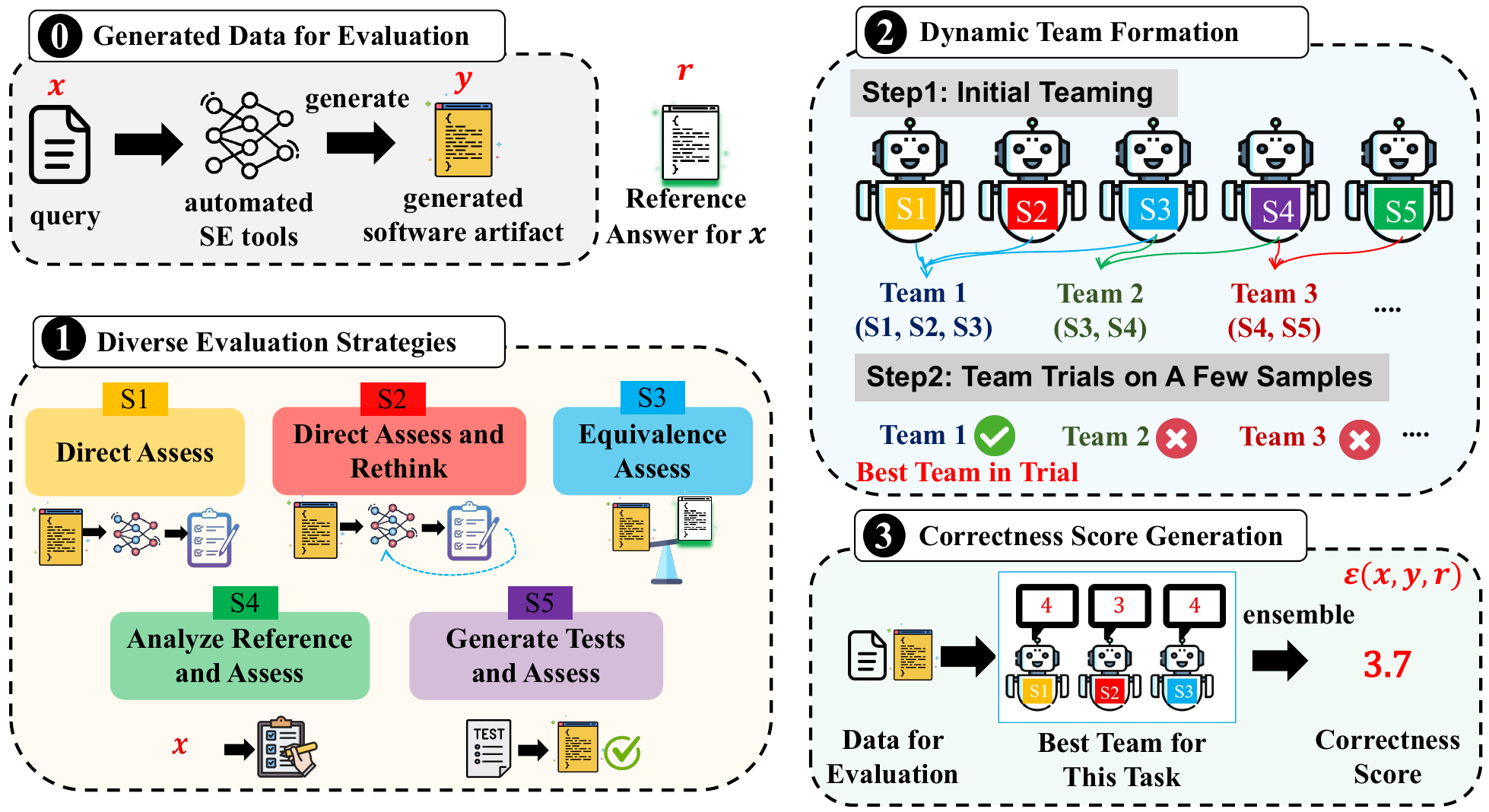}
    \vspace{-0.1cm}
    \caption{Overview of \TOOL.} 
    \label{fig:approach}
\vspace{-0.4cm}
\end{figure*}

\subsection{Problem Statement}
Automatic evaluation metrics aim to assess the quality of software artifacts generated by SE automation tools. 
In this work, \textbf{we focus specifically on the aspect of functional correctness for code-based artifacts}, which refers to the extent to which a generated software artifact fulfills the intended functional behavior described in the user requirement.
Correctness is a fundamental and indispensable attribute in many SE tasks. Without correctness, other desirable properties—such as efficiency or readability are rendered secondary.
For text–based artifacts, such as code summaries, functional correctness is not clearly defined. Therefore, we focus on the Content Adequacy aspect~\cite{SIDE}, which is more closely related to correctness than other aspects, e.g., fluency or conciseness. This Content Adequacy aspect assesses the extent to which the summary provides the information necessary to understand the code.

Formally, the task is defined as follows and illustrated in \circled{1} of Figure~\ref{fig:approach}. Let \( x \) denote a user's requirement (e.g., a natural language description of a task), and let \( y \) be a software artifact generated by an automated SE tool (e.g., an LLM-based code generator), intended to fulfill the requirement \( x \).  Let \( r \) be a reference solution that correctly fulfills the user requirement  \( x \).
For each generated software artifact \( y \), human annotators provide a correctness score, e.g., \( S\in\{0, 1, 2, 3, 4\} \), where 0 indicates a completely incorrect software artifact and 4 indicates a fully correct one. Our objective is to develop an \textit{automatic evaluation metric} \( \mathcal{E}(x, y, r) \) that can closely correlate with the human-provided correctness score \( S \). 
Our method supports both categorical values (e.g., 
\( S\in\{0, 1, 2, 3, 4\} \)) and continuous values. We use categorical values here because the human-annotated scores in the studied datasets follow this scheme. To adapt our method for continuous values, the only required modification is to adjust the prompt instructions so that the LLM generates continuous values instead of integers.

\subsection{Representative LLM-based Metric: ICE-Score}
\noindent The representative LLM-as-judge evaluation metric for code\label{sec:sota}, ICE-Score~\cite{ICE-Score}, prompts LLMs to directly assign evaluation scores to the generated software artifacts. Formally, it takes the requirement \( x \), the generated software artifact \( y \), and an optional reference solution \( r \), and inserts them into a predefined prompt, yielding \( Prompt(x, y, r) \). The LLM then generates a score based on this prompt: \( P = LLM(Prompt(x, y, r)) \). 
We showcase the ICE-Score's prompt for code generation:

\begin{tcolorbox}[colback=gray!5, colframe=black!70, title=Abstracted ICE-Score Prompt, width=\linewidth, boxrule=0.8pt, arc=2mm]
\textbf{[Task Description]} Your task is to rate the code snippet only on one metric ...

\textbf{[Evaluation Criteria]} Functional Correctness: Execution-based quality of the code snippet ...

\textbf{[Evaluation Steps]} 

1. Read the problem carefully;

2. Read the code snippet and compare it to the problem;

3. Assign a score for functional correctness ...

\textbf{[Data]} Problem: \textbf{x}, Code: \textbf{y},  Reference: \textbf{r}  (Optional)
\end{tcolorbox}

The core idea behind ICE-Score is to directly ``ask'' the LLM to assess the correctness of the generated code \( y \), as reflected in instructions like ``\textit{Your task is to rate the code snippet}'' and ``\textit{Assign a score for functional correctness.}'' This represents a straightforward strategy in using LLMs for correctness evaluation, which we refer to as the ``\textbf{Direct Assess}'' strategy. 
However, ICE-Score focuses solely on this strategy, leaving other potential strategies unexplored.

\subsection{\textbf{Motivation of \TOOL}}

\noindent\textbf{Motivation for Framework.}
Our work is inspired by the deliberative process of a legal tribunal composed of expert judges. In complex court cases, a panel of legal professionals is formed, with each member specializing in a specific domain such as criminal law, civil law, or international law. Each expert analyzes the case using their own legal framework, drawing on different reasoning approaches. The final verdict emerges through a process of collective deliberation, which benefits from the diversity of these expert perspectives. Such tribunals do not simply assemble all available experts; rather, they carefully select members whose expertise is most relevant to the case at hand, ensuring both efficiency and quality of deliberation.

Inspired by this legal setting, we identify two key limitations in existing evaluation approaches, such as ICE-Score. First, they rely on a single evaluator using a fixed strategy, which limits the diversity of perspectives. Second, they do not provide an effective mechanism for assembling a team of evaluators with complementary strengths. To address these limitations, our work introduces two key ideas at the framework level: \textbf{1) designing diverse evaluation strategies} that reflect different modes of reasoning, ensuring that varied analytical perspectives are brought into the assessment process; \textbf{2) introducing a lightweight team assembly mechanism} that selects an effective combination of evaluation strategies.

\vspace{0.1cm}
\noindent\textbf{Motivation for Correctness Evaluation Strategies.} 
When designing detailed strategies for evaluating correctness, we drew inspiration from a common and relevant academic task: grading exam papers, such as those in an algorithms course. In practice, different graders can apply different approaches to assess correctness. Some graders make a direct judgment about whether an answer is correct and assign a score accordingly. Others may revisit their initial evaluations to check for potential oversights. To improve efficiency, some graders might compare a student's solution to a reference answer and judge whether they are functionally or semantically equivalent. For more complex or subjective problems, graders may identify key evaluation criteria first and then verify whether each criterion is met. Inspired by this process, we propose several distinct perspectives for evaluating the correctness of software artifacts:

\begin{itemize} [left=2pt]
    \item [-] Assessment with justification and self-reflection, where an LLM provides a judgment, explains its reasons, and then re-evaluates the validity of those reasons;
    \item [-] Equivalence-based comparison, where an LLM determines whether the software artifact under evaluation is semantically or functionally equivalent to a reference;
    \item [-] Criteria-based evaluation, where an LLM first identifies the necessary conditions for correctness and then checks whether each condition is satisfied;
\end{itemize}
Lastly, recognizing the widespread use of test cases for verifying code correctness, we propose a test-based perspective. Since many of our evaluation datasets provide only a reference solution without test cases, we first generate test cases and then assess whether the code passes them. Each perspective above brings a distinct way of reasoning about correctness.

%% file: sections/3_approach.tex
\section{\label{overview}Our Approach}

Figure~\ref{fig:approach} presents the framework of SE-Jury, consisting of
three main stages:

\noindent\textbf{Stage 1: Diverse Evaluation Strategies (\circled{1} of Figure~\ref{fig:approach}).}
Given a requirement \( x \), a generated software artifact \( y \), and a reference solution \( r \), this part defines five distinct correctness evaluation strategies to assess the correctness of the generated software artifact \( y \) from diverse perspectives.

\vspace{0.1cm}
\noindent\textbf{Stage 2: Dynamic Team Formation (\circled{2} of Figure~\ref{fig:approach}).}
Given the five evaluation strategies, this stage aims to assemble an effective subset, referred to as a ``team'', from these strategies. This process enables the assembled team to adapt to varying dataset characteristics while reducing the costs of using LLMs.

\vspace{0.1cm}
\noindent\textbf{Stage 3: Correctness Score Generation (\circled{3} of Figure~\ref{fig:approach}).}
Once the team is determined, it is used to evaluate the correctness of data samples in the evaluation dataset, generating individual scores for each data sample. These individual scores are then aggregated to produce the final correctness score.

\subsection{Diverse Evaluation Strategies}

\vspace{0.1cm}
\noindent\textbf{Basics of Evaluation Strategy.} 
Following the prior work~\cite{ICE-Score}, our tool leverages the \textit{zero-shot capabilities} of LLMs, which means that we do not provide any human-annotated scores as input to LLMs. This choice reduces both manual labeling effort and LLM usage costs. For each evaluation data sample, we construct prompts containing the requirement \( x \), the generated software artifact \( y \), and the reference solution \( r \). These prompts are then fed into LLMs, which generate the correctness of  \( y \).

Each unique prompt design corresponds to a distinct evaluation strategy. By pairing a specific strategy's prompt with an LLM, we form an \emph{evaluator} that generates individual correctness scores in a zero-shot fashion. 
We introduce the prompt designs of five different evaluation strategies as follows:

\vspace{0.1cm}
\noindent\textbf{Strategy 1: Direct Assess.} 
Strategy 1 (S1) directly asks LLMs to assess the correctness of the generated software \( y \). 
For the prompt design of S1, see Section~\ref{sec:sota}. S1 has two variants: one without the reference solution and one with it. A team is considered to adopt S1 if it selects either variant.

\vspace{0.1cm}
\noindent\textbf{Strategy 2: Direct Assess and Rethink.}
Strategy 2 (S2) builds upon Strategy 1 (S1). In S1, the LLM directly provides a correctness score $\hat{s_1}$ for the generated software artifact \( y \), typically accompanied by a brief explanation justifying the assigned score. Inspired by the way humans often reflect on their initial judgments, S2 introduces a \textit{rethink} step. This step prompts the LLM to review both its previously assigned score and the reasoning behind it, and to consider whether any revision is necessary.
Specifically, the LLM is asked to critically re-evaluate the validity of its earlier explanation and adjust its score accordingly. For example, suppose in S1 the LLM assigns a low score to the generated software artifact $y$ due to a flaw it identifies (e.g., a reason $e$). During the rethink phase, if the LLM realizes that this reason $e$ is actually incorrect, it is encouraged to revise the score upward. 
Below, we present an example of the prompt in S2 for the code generation task.

\begin{tcolorbox}[colback=gray!5, colframe=black!70, title=Prompt of Strategy 2, width=\linewidth, boxrule=0.8pt, arc=2mm]

\textbf{$<$Response from Strategy 1: predicted score $\hat{s_1}$ and its reasons $e$$>$}

\textcolor{darkgreen}{\textit{`````` Prompt Segment Unique to Strategy 2 """}}

\textbf{[Task Description]} Your task is to recheck whether the reason and score are proper...

\textbf{[Evaluation Criteria]} 

1. If a bad reason about the code snippet is validated to be `False', increase the score...

2. If a good reason about the code snippet is validated to be `False', decrease the score...

3. If a reason is `True', do not change the score...

\textbf{[Evaluation Steps]}

1. Please only validate the previous score and reason.

2. Please reply with your adjusted score.

\textbf{[Data]} Problem: \textbf{x}, Code Snippet: \textbf{y},  Predicted Score from S1: \textbf{$\hat{s_1}$}, Reasons from S1: \textbf{e}
\end{tcolorbox}

\noindent
After the rethink step, the LLM will produce an adjusted score \( \hat{s_2} \) by either increasing, decreasing, or maintaining the original correctness score generated in S1 (i.e., $\hat{s_1}$).

\begin{tcolorbox}[colback=gray!5, colframe=black!70, title=Prompt of Strategy 3, width=\linewidth, boxrule=0.8pt, arc=2mm]

\textbf{[Task Description]} Given two code, your task is to assess whether they are semantically equivalent...

\textbf{[Evaluation Criteria]} Semantic Equivalence: To what extent do the two code produce the same behavior...

\textbf{[Evaluation Steps]}

1. Read and analyze both code versions carefully. Read the problem description too...

2. Compare their functionality, structure, and logic to determine if they yield the same output and behavior...

3. Assign a Semantic Equivalence score...

\textbf{[Data]} Problem: \textbf{x}, Code: \textbf{y}, Reference: \textbf{r}
\end{tcolorbox}

\begin{figure*}[t]
    \centering
    \includegraphics[width=0.78\textwidth]{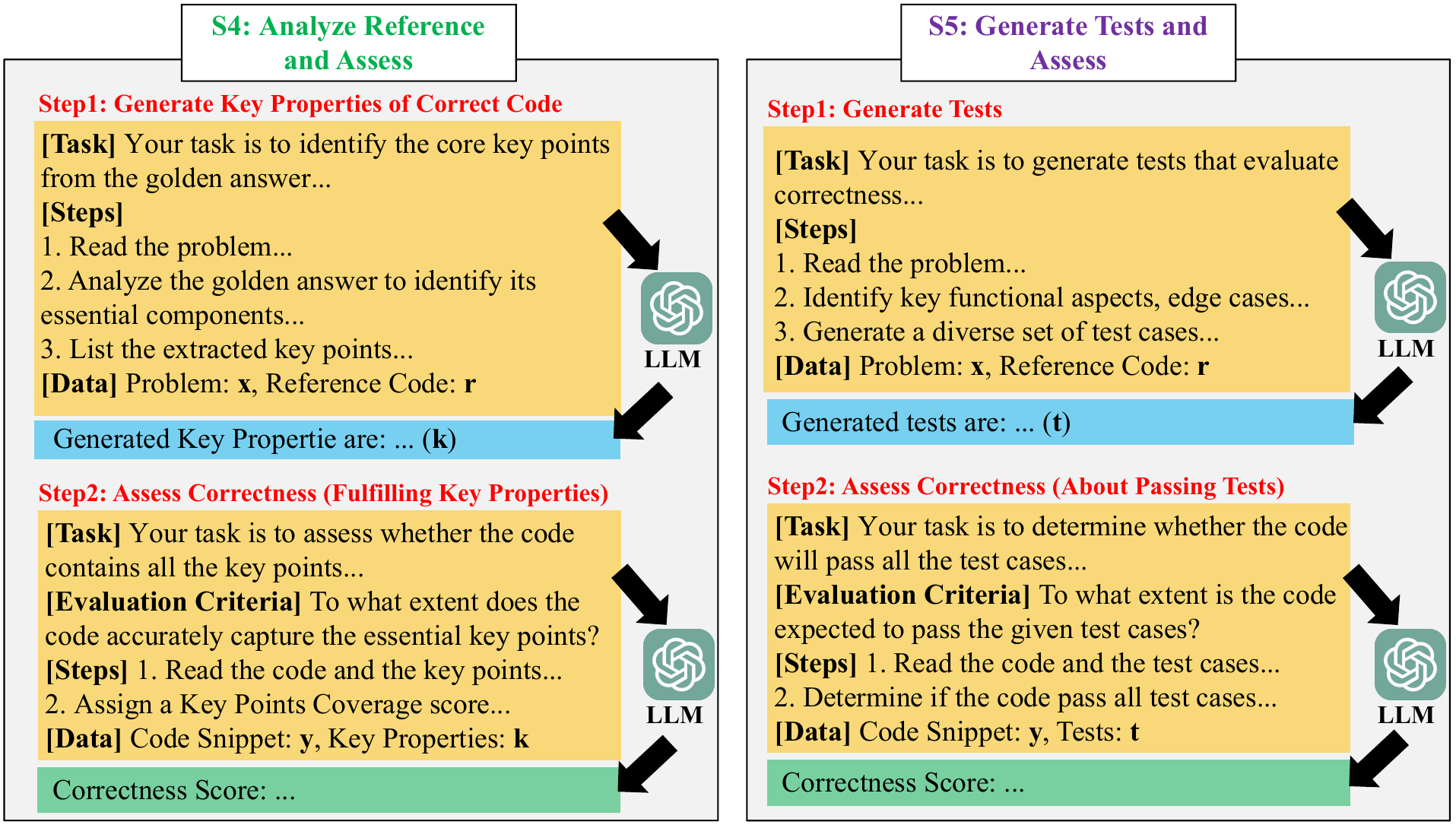}
    \vspace{-0.5em}
    \caption{Prompt Designs of Strategy 4 and Strategy 5.}
    \label{fig:p4_p5}
\vspace{-1.0em}
\end{figure*}

\noindent\textbf{Strategy 3: Equivalence Assess.}
Strategy 3 (S3) adopts a different approach from S1 and S2. Since the reference solution, \( r \), can correctly satisfy the user requirement \( x \), we can assess the correctness of \( y \) by evaluating its equivalence to \( r \).
The underlying idea is that if \( y \) and \( r \) are semantically or functionally equivalent, then it is highly likely that \( y \) also meets the original requirement \( x \). Therefore, rather than reasoning directly on the correctness of \( y \), the LLM focuses on comparing \( y \) and \( r \), making this strategy an equivalence-based evaluation strategy. We present an example of the prompt used in S3 for the code generation task above.

\vspace{0.1cm}
\noindent\textbf{Strategy 4: Analyze Reference and Assess.}
Strategy 4 (S4) adopts an analytical approach. 
Figure~\ref{fig:p4_p5} illustrates the prompt design for S4. This strategy involves two steps. First, the LLM identifies the critical properties of \( r \) that make it a correct solution for \( x \). In the second step, the LLM checks whether \( y \) preserves those core properties. If the LLM determines that \( y \) aligns with the reference solution's key characteristics, it considers \( y \) to be correct.

\vspace{0.1cm}
\noindent\textbf{Strategy 5: Generate Tests and Assess.}
Strategy 5 (S5) introduces a different evaluation strategy. The core idea is straightforward: when the generated artifact \( y \) is a code snippet or a code change, test cases serve as an effective means for assessing its correctness. 
Figure~\ref{fig:p4_p5} illustrates the prompt design for S5, which consists of two steps. In the first step, we prompt the LLM to generate test cases based on the user requirement \( x \) and the reference code \( r \). In the second step, we provide the generated software artifact \( y \) along with the previously generated test cases as input to the LLM. The LLM is then asked to evaluate whether \( y \) can pass all the generated test cases and, based on this evaluation, assign a correctness score.
We generate test cases using the reference code, as using the generated code directly could propagate any existing bugs into the test cases. By relying on the known reference code, the LLM can produce more reliable test cases.

\vspace{0.1cm}
\noindent
Please note that, due to space limitations, we primarily present the prompt design at a high level for the code generation task, and the prompts shown are simplified illustrative versions.

\vspace{0.1cm}
\noindent\textbf{Turning Strategies into Evaluators.}
With all strategies defined, we can now pair each strategy with a specific LLM to construct a set of evaluators. Each evaluator produces an independent correctness score according to its respective evaluation strategy.
In addition, as the target human score ranges vary across datasets such as CoNaLa~\cite{DBLP:journals/jss/EvtikhievBSB23} using a 0–4 scale, and Summary-Assess~\cite{DBLP:conf/sigsoft/RoyFA21} using 1–5, we standardize the output range across all strategies to ensure consistency. To do this, we include an instruction in each prompt that constrains the LLM to output a score within the 0–100 range.

\subsection{Dynamic Team Formation}
Dynamic team selection aims to choose the most suitable team for each dataset, helping to reduce costs by avoiding the use of all strategies on every task and excluding those that are less effective for specific tasks. For example, since test cases cannot be generated for code comments, Strategy S5 is not suitable for evaluating the correctness of code comments.
Building on this insight, this component dynamically assembles an effective team from the available 5 strategies, tailoring the combination to each dataset.

\vspace{0.1cm}
\noindent\textbf{Initial Teaming.}  
Let \( \mathcal{S} = \{S_{1}, S_{2}, S_3, S_4, S_5\} \) denote the set of strategies we can choose from. We leverage LLMs to automatically obtain correctness scores, allowing us to explore a broad space of strategy combinations. Specifically, we consider all combinations that include at least two distinct strategies and require S1, since it represents the most fundamental strategy that directly reflects the original task.
We denote these combinations as \( \mathcal{T} = \{T_1, \ldots, T_{k}\} \), where $k$ is the number of possible combinations.

\vspace{0.1cm}
\noindent\textbf{Team Trials on A Few Annotated Samples.}  
To identify the best team from $\mathcal{T} = \{T_1, \ldots, T_{k}\}$, we utilize a small set of annotated examples from the evaluation dataset. We randomly sample 20 instances, assuming their ground truth correctness scores are available, as annotating this number is feasible for a human developer. Each team generates predicted scores for these samples, and we measure their alignment with the ground truth 
using Kendall's $\tau$ coefficient and Spearman's $r_s$. The team with the highest correlation is selected as the best team.

Although they may appear similar, dynamic team selection serves a different purpose and operates under different constraints compared to Hyperparameter Tuning (HP). HP is applied during training and requires a large amount of labeled data. In contrast, our method selects the ensemble configuration and only uses 20 annotated samples.

\subsection{Final Correctness Score Generation} 
For illustration purposes, suppose the selected team \( T_i \) comprises strategies \( S_1 \), \( S_2 \), and \( S_3 \). 
Please note that \( T_i = (S_1, S_2, S_3)\) is just an example.

\vspace{0.1cm}
\noindent\textbf{Individual Score Prediction.}~\label{final_score_generation}
Given the selected team \( T_i = (S_1, S_2, S_3) \), we apply this team to the remaining samples in the dataset \( \mathcal{D} = \{d_1, d_2, \ldots, d_k\} \), excluding the 20 randomly sampled instances used for team selection.
We exclude these 20 instances from the evaluation dataset to avoid bias, as their ground truth scores were involved in the selection process.

Each sample \( d_i \in \mathcal{D} \) is represented as a tuple \( (x_i, y_i, r_i) \). Strategies in the selected team independently generate correctness scores for each sample, resulting in three individual scores: \( s_1 \), \( s_2 \), and \( s_3 \), corresponding to \( S_1 \), \( S_2 \), and \( S_3 \), respectively.

\vspace{0.1cm}
\noindent\textbf{Score Ensembling.}
To generate the final score for each sample, we aggregate the individual scores from the team members using a simple averaging ensembling strategy. The final predicted score $\hat{s}$ is computed as: $\hat{s} = \frac{s_1 + s_2 + s_3}{3}$.

\vspace{0.1cm}
\noindent\textbf{Mapping Score to Target Scale.}
The predicted score $\hat{s}$ is initially on a 0--100 scale. However, human-annotated scores in different datasets may use different scales (e.g., 1--5). To ensure compatibility with the evaluation criteria, we apply a linear transformation~\cite{cullen2012matrices}. 
For example, for datasets where the human-annotated scores follow a 1--5 scale, we map the predicted score as follows: $\mathcal{E}(x, y, r) = \frac{\hat{s}}{100} \times 4 +1$.

%% file: sections/4_setup.tex
\section{Experimental Setup\label{experiment}}
In this section, we introduce the datasets used in our experiments, the baseline methods for comparison, and the evaluation methodology to assess the effectiveness of our proposed metric. We also outline the implementation details and define the key research questions.

\subsection{Datasets}
We evaluate \TOOL on three popular SE tasks: code generation, automated program repair, and code summarization. The primary goal is to assess how well \TOOL's results align with human evaluation results. Therefore, we have selected evaluation datasets that include human evaluation scores. Moreover, all selected datasets contain the reference solutions. The selected datasets are:

\begin{itemize}[left=2pt]

\item \textbf{CoNaLa}~\cite{DBLP:conf/msr/YinDCVN08} is a Python \textit{Code Generation} benchmark sourced from StackOverflow. We selected CoNaLa because Evtikhiev et al.\cite{DBLP:journals/jss/EvtikhievBSB23} provided human evaluation scores for code generated by various automated code generation tools addressing these CoNaLa problems.

\item \textbf{Card2Code}~\cite{Card2Code} is a Python \textit{Code Generation} benchmark derived from the collectible trading card game Hearthstone. 
We selected Card2Code due to its inclusion in~\cite{DBLP:journals/jss/EvtikhievBSB23}.

\item \textbf{APR-Assess}~\cite{APR-assess} is a human-annotated dataset for \textit{Automated Program Repair (APR)}, involving the generation of patches (i.e., code changes) to fix identified bugs. It consists of patches generated by program repair tools, each manually evaluated for correctness. 
Although the APR data contains tests, tests in real-world projects can be weak~\cite{ye2021automated,qi2015analysis}. As a result, an APR-generated patch passing all the tests could still be faulty upon developer review~\cite{ye2021automated}. In this study, we use the APR-Assess dataset~\cite{APR-assess}, where 35 professional developers manually reviewed test-passing patches to determine their true correctness. These labels are considered more reliable than test outcomes, as they are assigned by experts with full knowledge of the test results.

\item \textbf{Summary-Assess}~\cite{DBLP:conf/sigsoft/RoyFA21} is a human-annotated dataset for \textit{Code Summarization}, which focuses on generating accurate descriptions for given code snippets. It is based on a publicly available Java code summarization dataset~\cite{leclair2019recommendations}. Human annotators evaluate various aspects of each summary, including conciseness, fluency, and content adequacy.
We use the content adequacy results as the ground truth labels.

\end{itemize}

\subsection{Selected Baselines}

\noindent\textbf{Match-based Metrics.} 
We select 7 widely used match-based metrics as baselines, all of which evaluate the similarity between generated content and ground-truth answers. These include four from the natural language processing domain: BLEU~\cite{BLEU}, ROUGE-L~\cite{lin2004rouge}, ChrF++~\cite{DBLP:conf/wmt/Popovic17}, METEOR~\cite{DBLP:conf/acl/BanerjeeL05}, and three designed for SE: CodeBLEU~\cite{DBLP:journals/corr/abs-2009-10297}, RUBY~\cite{DBLP:conf/iwpc/TranTNNN19}, and CrystalBLEU~\cite{DBLP:conf/kbse/EghbaliP22}. Among them, CrystalBLEU~\cite{DBLP:conf/kbse/EghbaliP22} is the SOTA match-based metric designed to measure code similarity.

\vspace{0.1cm}
\noindent\textbf{Embedding-based Metric.} 
We choose four popular embedding-based metrics as baselines:
MoverScore~\cite{DBLP:conf/emnlp/ZhaoPLGME19}, BERTScore~\cite{DBLP:conf/iclr/ZhangKWWA20}, CodeBERTScore, and SIDE~\cite{SIDE}. 
Among these, SIDE~\cite{SIDE} is the state-of-the-art metric, though it is specifically designed for the code summarization task.

\vspace{0.1cm}
\noindent\textbf{LLM-as-judge Metrics.} 
We select three LLM-as-judge metrics as baselines.
\textit{Vanilla LLM} refers to the default LLM used with a straightforward prompt, without employing the specialized strategies proposed in this work. Specifically, we provide the LLM with a simple instruction:``Please assign a correctness score to the given input data.''
ICE-Score~\cite{ICE-Score} is the leading LLM-as-judge method for code evaluation. It extends the recent LLM-as-judge approach for text, G-Eval~\cite{DBLP:conf/emnlp/LiuIXWXZ23}, with adaptations for code evaluation. ICE-Score prompts the LLM to generate a correctness score based on pre-defined evaluation criteria. To enable ICE-Score to support automated program repair and code summarization tasks, we adapted it (originally designed for code evaluation) by implementing similarly styled prompts. 
The adapted ICE-Score prompt retains the same core modules as the original framework: (1) task description, (2) evaluation criteria, (3) evaluation steps, and (4) data placeholders, with only necessary modifications (e.g., task definitions, evaluation criteria explanations). For all datasets, we evaluated two variants of ICE-Score: one using reference answers and one without, and reported the better-performing variant on average. CodeJudge~\cite{tong2024codejudge} is a state-of-the-art LLM-based evaluator of code correctness, designed to assess the functional correctness of generated code.

\subsection{Effectiveness Evaluation}
We use two evaluation approaches to assess the effectiveness.

\vspace{0.1cm}
\noindent\textbf{Statistical Correlations.} 
We follow prior studies~\cite{ICE-Score,tong2024codejudge} to employ two statistical correlation metrics, including \textit{Kendall's $\tau$ coefficient} and \textit{Spearman's $r_s$}. They are widely used to measure the statistical correlation between evaluation results produced by automatic evaluation metrics and the ground truth. 
Specifically, Kendall's $\tau$ coefficient~\cite{kendall1938new} measures the ordinal association between two data, and Spearman $r_s$~\cite{spearman1961proof} is a measure of rank correlation.
Note that Pearson’s correlation is not employed because it is a parametric measure that assumes a linear relationship between variables, an assumption that may not be satisfied in our tasks.
For ease of comparison, we report the average of the Kendall's and Spearman's correlation coefficients.

\vspace{0.1cm}
\noindent\textbf{Statistical Agreements.}
We also evaluate the statistical agreement between our tool's results and human evaluation scores. Specifically, we use \textit{Cohen's Kappa score}~\cite{mchugh2012interrater}, a statistical measure that assesses the agreement between two raters who independently classify items into categories.

\input{sections/tables/main_results}

\subsection{Implementation Details}
We evaluate the effectiveness of \TOOL using the OpenAI GPT-4o mini model (i.e., \textit{gpt-4o-mini-2024-07-18})~\cite{openai_gpt4o_mini} as the backbone. 
For reasons of scalability and cost-effectiveness, we selected GPT-4o-mini as the primary underlying LLM. 
As a lightweight version of GPT-4o, GPT-4o-mini costs only US\$0.15 per million input tokens, which is approximately 6\% of the cost of GPT-4o, 1.5\% of GPT-4-turbo, and 30\% of GPT-3.5-turbo.
We set the temperature to 0 to reduce the impact of randomness in the LLM on the results.

\subsection{Research Questions}
Our work aims to answer three Research Questions (RQs). 
\begin{itemize}[leftmargin=*]
\item \textbf{RQ1: How well does \TOOL correlate with human judgment compared to baseline methods?} 
In RQ1, we investigate whether \TOOL generates evaluation results that more closely correlate with human judgment compared to baseline evaluation metrics.

\item \textbf{RQ2: How does the agreement between \TOOL and human evaluators compare to the agreement among humans?}
In RQ2, we quantify the gap between human-tool agreement and human-human agreement to assess how closely \TOOL can replace human evaluators.

\item \textbf{RQ3: How do the key design components of \TOOL impact its effectiveness?}
We conduct an ablation study to assess the contributions of the main modules of \TOOL. We also analyze the impact of different design choices.

\item 
\textbf{RQ4: How well does \TOOL correlate with test case execution outcomes?}
In RQ4, we investigate the generalizability of SE-Jury from human-annotated labels to labels derived from test execution. 
\end{itemize}

%% file: sections/tables/main_results.tex
\begin{table*}[t]
    \centering
    \caption{Experimental results for correlation with human scores. The highest correlation is highlighted in bold, and the best-performing baseline is underlined. $\tau$: Kendall's coefficient, and $r_s$: Spearman's coefficient.}
     \begin{tabular}{l|ccc|ccc|ccc|ccc|>{\columncolor{red!10}}c}
    \hline
    & \multicolumn{3}{c|}{\textbf{CoNaLa}} & \multicolumn{3}{c|}{\textbf{Card2Code}} & \multicolumn{3}{c|}{\textbf{APR-Assess}} & \multicolumn{3}{c|}{\textbf{Summary-Assess}} & \multicolumn{1}{c}{\textbf{Average}}\\
    
    \textbf{Metrics}  & $\tau$  & $r_s$ & avg & $\tau$ &  $r_s$ & avg & $\tau$ &  $r_s$ & avg & $\tau$ &  $r_s$ & avg & \cellcolor{red!10}\textbf{Tasks} \\ 
    \thickcline{1-14}
    \rowcolor{gray!18} \multicolumn{13}{c|}{\textbf{Existing Metrics}} & \cellcolor{red!10}\\ 
  \multicolumn{1}{l|}{BLEU} & 29.4 & 32.6 & \multicolumn{1}{c|}{31.0} & 43.8 & 54.3 & \multicolumn{1}{c|}{49.0} & 24.3 & 29.6 & 26.9 & 13.4 & 15.3 & 14.3 & \cellcolor{red!10}30.4 \\
\multicolumn{1}{l|}{ROUGE-L} & 44.0 & 51.0 & \multicolumn{1}{c|}{47.5} & 52.6 & 65.0 & \multicolumn{1}{c|}{58.8} & 20.1 & 24.6 & 22.4 & 17.6 & 21.7 & 19.6 & \cellcolor{red!10}37.1 \\
\multicolumn{1}{l|}{METEOR} & 38.0 & 44.3 & \multicolumn{1}{c|}{41.2} & 65.4 &  \underline{79.6} & \multicolumn{1}{c|}{\underline{72.5}} & 40.5 & \underline{49.4} & \underline{45.0} & 17.8 & 22.0 & 19.9 & \cellcolor{red!10}44.7 \\
\multicolumn{1}{l|}{Chrf++} & 47.3 & 54.4 & \multicolumn{1}{c|}{50.9} & 64.0 & 78.3 & \multicolumn{1}{c|}{71.1} & 23.1 & 28.2 & 25.7 & 19.0 & 23.2 & 21.1 & \cellcolor{red!10}42.2 \\
\multicolumn{1}{l|}{CodeBLEU} & 22.4 & 25.3 & \multicolumn{1}{c|}{23.9} & 40.9 & 51.3 & \multicolumn{1}{c|}{46.1} & 19.7 & 24.0 & 21.8 & 13.3 & 16.0 & 14.6 & \cellcolor{red!10}26.7 \\
\multicolumn{1}{l|}{RUBY} & 36.5 & 42.9 & \multicolumn{1}{c|}{39.7} & 65.4 & 78.2 & \multicolumn{1}{c|}{71.8} & 10.4 &  12.7 & 11.6 & 17.0 & 20.9 & 18.9 & \cellcolor{red!10}35.5 \\
\multicolumn{1}{l|}{CrystalBLEU} & 26.7 & 29.5 & \multicolumn{1}{c|}{28.1} & 37.8 & 47.6 & \multicolumn{1}{c|}{42.7} & 28.7 & 35.0 &31.8 & 12.8 & 14.6 & 13.7 & \cellcolor{red!10}29.1 \\
\multicolumn{1}{l|}{MoverScore} & 40.0 &  44.0 & \multicolumn{1}{c|}{42.0} & 64.7 &  79.0 & \multicolumn{1}{c|}{71.8} & 17.7 &  21.6 & 19.6 & 15.8 &  18.3 & 17.0 & \cellcolor{red!10}37.7 \\
\multicolumn{1}{l|}{BERTScore} & 43.8  & 48.0 & \multicolumn{1}{c|}{45.9} & 55.8 &  69.1 & \multicolumn{1}{c|}{62.5} & 1.3  & 1.6 & 1.5 & 21.6 &  24.0 & 22.8 & \cellcolor{red!10}33.2 \\
\multicolumn{1}{l|}{CodeBERTScore} & 42.1 & 46.6 & \multicolumn{1}{c|}{44.3} & 58.0 & 72.2 & \multicolumn{1}{c|}{65.1} & 5.5 & 6.7 & 6.1 & 14.8 & 17.0 & 15.9 & \cellcolor{red!10}32.9 \\
\multicolumn{1}{l|}{SIDE} & - & - & \multicolumn{1}{c|}{-} & - &- & \multicolumn{1}{c|}{-} & - & - & - & 23.8 & 28.2 & 26.0 & \cellcolor{red!10}- \\
\multicolumn{1}{l|}{Vanilla LLM} & 42.3 & 48.9 & \multicolumn{1}{c|}{45.6} & \underline{66.1} &  74.6 & \multicolumn{1}{c|}{70.4} & \underline{42.6}  & 44.5 & 43.5& \underline{31.1}  & \underline{36.7} & \underline{33.9} & \cellcolor{red!10}48.4 \\
\multicolumn{1}{l|}{ICE-Score} & \underline{52.1}  & \underline{58.2} & \multicolumn{1}{c|}{\underline{55.2}} & 62.7 & 70.2 & \multicolumn{1}{c|}{66.5} & 42.1 &  44.9 & 43.5 & 30.1 & 36.0 & 33.1 & \cellcolor{red!10}\underline{49.6} \\
    \thickcline{1-14}
    \rowcolor{green!5} \multicolumn{13}{c|}{\textbf{Our LLM-as-Ensemble-Judge Metric}} & \cellcolor{red!10} \\ 
    \textbf{\TOOL}     & \textbf{60.0} & \textbf{66.9} & \multicolumn{1}{c|}{\textbf{63.5}} & \textbf{76.9} & \textbf{83.7} & \multicolumn{1}{c|}{\textbf{80.3}} & \textbf{76.2} & \textbf{76.2} & \textbf{76.2} & \textbf{33.2}  & \textbf{41.5} & \textbf{37.3} & \cellcolor{red!10}\textbf{64.3} \\
    \multicolumn{1}{l|}{w/o Team Select (Merge All)} & 59.3 & 66.5 & \multicolumn{1}{c|}{62.9} & 73.4 &  80.4 & \multicolumn{1}{c|}{76.9} & 73.6 & 73.6 &73.6 & 29.8 & 38.2 & 34.0 & \cellcolor{red!10}61.9 \\
    \textit{w/o Team \& Strategies}     & 42.3 & 48.9 & \multicolumn{1}{c|}{45.6} & 66.1 & 74.6 & \multicolumn{1}{c|}{70.4} & 42.6 & 44.5 & 43.5 & 31.1 & 36.7 & 33.9 & \cellcolor{red!10}48.4  \\ 
    \hline
\end{tabular}
    \label{tab:RQ1_res}
\end{table*}

%% file: sections/5_results.tex
\section{Experimental Results\label{result}}

\input{sections/5_RQ1}

\input{sections/5_RQ2}

\input{sections/5_RQ3}

\input{sections/5_RQ4}

%% file: sections/5_RQ1.tex
\subsection{RQ1: Correlation with Human Scores}

In RQ1, we evaluate the correlation between \TOOL's scores and human-annotated scores.
Table~\ref{tab:RQ1_res} shows how well \TOOL's scores correlate with human judgments across four human-annotated datasets. For the results of each dataset, the first two columns report statistical correlation metrics: Kendall’s $\tau$ and Spearman’s $r_s$, respectively.

\textbf{\TOOL achieves the highest alignment with human evaluations, consistently and significantly outperforming all baseline methods.}
As shown in Table~\ref{tab:RQ1_res}, \TOOL consistently outperforms all baselines by an average of 29.6\%–140.8\% across all tasks, based on the mean of two statistical correlation metrics. 
Among the baselines, ICE-Score performs the best. 
For individual tasks, \TOOL exceeds the best-performing baseline ICE-Score by 17.9\% 
on the code generation task (CoNaLa and Card2Code), 75.2\% on the automated program repair task (APR-Assess), and 12.7\% on the code summarization task (Summary-Assess).
The results indicate that \TOOL generalizes well across diverse types of software artifacts: code changes in APR-Assess, code snippets in CoNaLa and Card2Code, and code comments in Summary-Assess.
Moreover, we perform the Wilcoxon signed-rank test~\cite{conover1999practical} at the 95\% significance level (i.e., p-value $<$ 0.05) between \TOOL and each baseline. 
The Wilcoxon signed-rank test was applied to the paired performance differences between SE-Jury and each baseline over all studied datasets together. Specifically, the performance was measured as the difference between a tool’s output score and the human score. To account for multiple comparisons, the Bonferroni correction~\cite{bonferroni1936teoria} was applied.
The results show that p-values are below 0.05, indicating that \TOOL significantly outperforms baselines.

\textbf{\TOOL achieves strong alignment with human judgments in code generation and automated program repair.}
We follow prior work to interpret correlation scores: Kendall’s $\tau \ge 0.5$~\cite{bachmann2010process} and Spearman’s $r_s \ge 0.6$~\cite{akoglu2018user} are considered strong correlations.
Based on these thresholds, \TOOL shows strong alignment with human evaluation for code generation (CoNaLa and Card2Code) and automated program repair (APR-Assess).
While \TOOL's performance on code summarization (Summary-Assess) is relatively lower compared to other tasks, it still substantially outperforms competing methods.

\vspace{0.3cm}
\noindent
\begin{tcolorbox} [boxrule=0.8pt,
                top=0.2pt,
                  bottom=0.2pt]
    \textbf{Answer to RQ1}: 
\TOOL achieves the highest alignment with human evaluations, consistently and substantially outperforming all baselines by 29.6\%–140.8\% on average. 
For all correlation metrics, \TOOL shows strong alignment with human evaluations except Summary-Assess, indicating its alignment with humans.

\end{tcolorbox}

%% file: sections/5_RQ2.tex
\subsection{RQ2: Human-Tool \& Human-Human Agreement}

In RQ2, we assess how closely \TOOL aligns with individual annotators relative to human–human agreement.

\begin{figure}[b] 
    \centering
    \vspace{-0.5cm}
    \includegraphics[width=9cm]{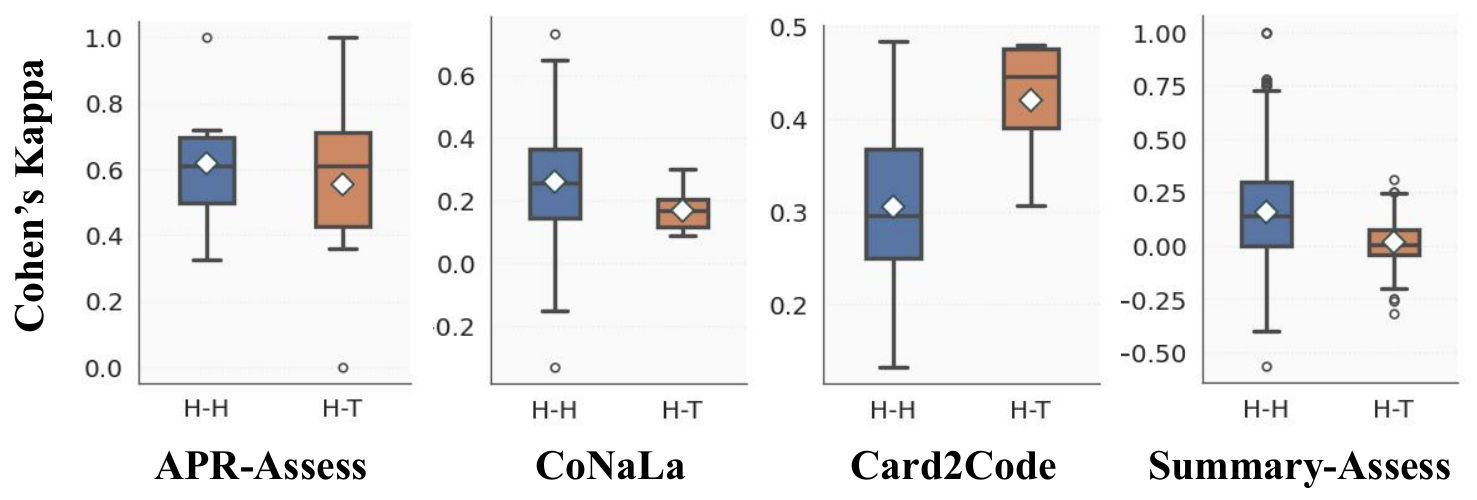} 
    \vspace{-0.5cm}
    \caption{Agreements between human developers (``H-H'' and highlighted in blue) and agreements between \TOOL and humans (``H-T'' and highlighted in orange).} 
    \label{fig:rq2}
\end{figure}

\vspace{0.1cm}
\noindent\textbf{Setup.}
It is important to note that the datasets we study, i.e., CoNaLa, Card2Code, APR-Assess, and Summary-Assess, are all annotated by multiple human developers. 
In RQ1, we used the aggregated human score as the ground truth. This score is obtained by combining the individual ratings from different annotators. For example, in the CoNaLa dataset, Evtikhiev et al. adopted the M-MSR algorithm~\cite{ma2020adversarial} to aggregate multiple human grades into a single aggregated human score~\cite{DBLP:journals/jss/EvtikhievBSB23}.
In contrast, \textit{RQ2 uses individual human annotations as the ground truth} to evaluate \TOOL. 
Our objective is to measure the gap between the agreement levels of \TOOL with human annotators (human-tool agreement) and the agreement among human annotators themselves (human-human agreement). 
If its agreement with individual annotators matches the level of agreement among humans themselves, it indicates that \TOOL could serve as a reliable surrogate for human evaluation.
Specifically, we group human annotators into pairs and compute Cohen’s Kappa~\cite{mchugh2012interrater} scores to quantify their agreement levels. 
Moreover, we pair \TOOL with each human annotator and compute Cohen’s Kappa scores on all human-tool pairs.

\input{sections/tables/impact}

\vspace{0.1cm}
\noindent\textbf{Results.} Figure~\ref{fig:rq2} shows a box plot illustrating the agreement levels between human developers (labeled as ``H-H'' in the figure and highlighted in blue) and between humans and our tool (labeled as ``H-T'' and highlighted in orange), measured by Cohen’s Kappa scores. The middle line of each box represents the median, and the white diamond represents the mean. A smaller box indicates more stable results and smaller variance.

\textbf{In the automated program repair task, \TOOL achieves agreement levels with human annotators that are comparable to the agreement among human annotators.}
Figure~\ref{fig:rq2} shows that \TOOL exhibits comparable median and mean agreement with human annotators, indicating it has the potential to serve as a reliable substitute for humans in this task.

\textbf{In the code generation task, \TOOL achieves agreement levels that are close to human-human agreement.}
Specifically, on CoNaLa, \TOOL attains a score of 16.7, lower than the human–human average of 25.7. In contrast, on Card2Code, it achieves a higher average Cohen’s Kappa score of 41.9, surpassing the human–human agreement score of 30.5.
Overall, \TOOL demonstrates performance close to that of human evaluators, making it a promising choice for code generation evaluation.

\textbf{However, in the code summarization task, there remains a substantial gap between the agreement of \TOOL and human annotators compared to human-human agreement.}
On the Summary-Assess dataset, \TOOL achieves an average Cohen’s Kappa score of 1.9, which is substantially lower than the human–human average of 15.5.
This suggests that \TOOL is not yet a viable replacement for human evaluators in code summarization.
Nonetheless, as shown in Table~\ref{tab:RQ1_res}, \TOOL remains the best-performing automatic evaluation metric in the code summarization task, highlighting our progress.

\vspace{0.3cm}
\noindent
\begin{tcolorbox} [boxrule=0.8pt,
                top=0.2pt,
                  bottom=0.2pt]
    \textbf{Answer to RQ2}: 
    Compared to inter-human agreement levels, \TOOL achieves comparable agreement with human annotators in the automated program repair task and approaches human-level agreement in the code generation task, indicating that \TOOL holds promise as a substitute for human evaluators in these tasks.
\end{tcolorbox}

%% file: sections/tables/impact.tex
\begin{table*}[t]
    \centering
    \caption{Performance of \TOOL with different LLMs, individual evaluation strategies, and the underlying LLM alone.
    } 
     \begin{tabular}{l|ccc|ccc|ccc|ccc|>{\columncolor{white!10}}c}
    \hline
    & \multicolumn{3}{c|}{\textbf{CoNaLa}} & \multicolumn{3}{c|}{\textbf{Card2Code}} & \multicolumn{3}{c|}{\textbf{APR-Assess}} & \multicolumn{3}{c|}{\textbf{Summary-Assess}} & \multicolumn{1}{c}{\textbf{Average}}\\
    
    \textbf{Metrics}  & $\tau$  & $r_s$ & avg & $\tau$ &  $r_s$ & avg & $\tau$ &  $r_s$ & avg & $\tau$ &  $r_s$ & avg & \cellcolor{white!10}\textbf{Tasks} \\ 
    \thickcline{1-14}
    \rowcolor{green!8} \multicolumn{14}{c}{\textbf{Performance of Using Different Underlying LLMs}} \\ 
    \textbf{\TOOL(GPT-4o-mini)}     & 60.0 & 66.9 & \multicolumn{1}{c|}{63.5} & 76.9 & 83.7 & \multicolumn{1}{c|}{80.3} & \textbf{76.2} & \textbf{76.2} & \textbf{76.2} & \textbf{33.2}  & \textbf{41.5} & \textbf{37.3} & \cellcolor{white!10}64.3 \\
     \textbf{\TOOL(DeepSeek-Chat)}     & \textbf{62.3} & \textbf{72.6} & \multicolumn{1}{c|}{\textbf{67.4}} & \textbf{80.5} & \textbf{86.7} & \multicolumn{1}{c|}{\textbf{83.6}} & 74.2 & 74.2 & 74.2 & 31.2  & 37.2 & 34.2 & \textbf{64.9} \\
    \hline

     \rowcolor{Cyan!10} \multicolumn{14}{c}{\textbf{Performance of Individual Strategies (with GPT-4o-mini)}} \\ 
\textbf{Team Selected by \TOOL}    & \multicolumn{3}{c}{S1+S4} & \multicolumn{3}{c}{S1+S2+S3} & \multicolumn{3}{c}{S1+S4} & \multicolumn{3}{c}{S1+S2} & - \\
\hdashline
    
    \textbf{S1}     & 56.9 & 63.3 & \multicolumn{1}{c|}{60.1} & 56.7 & 62.8 & \multicolumn{1}{c|}{59.8} & 68.6 & 68.6 & 68.6 & 33.2  & 39.0 & 36.1 & \cellcolor{white!10}56.2 \\
    \textbf{S2}     & 52.7 & 60.2 & \multicolumn{1}{c|}{56.4} & 72.5 & 80.2 & \multicolumn{1}{c|}{76.4} & 63.1 & 63.1 & 63.1 & 31.9  & 37.8 & 34.9 & \cellcolor{white!10}57.7 \\
    \textbf{S3}     & 52.3 & 59.0 & \multicolumn{1}{c|}{55.6} & 76.1 & 82.7 & \multicolumn{1}{c|}{79.4} & 76.9 & 76.9 & 76.9 & 19.5  & 23.2 & 21.4 & \cellcolor{white!10}\textbf{58.3} \\
    \textbf{S4}     & 45.5 & 50.3 & \multicolumn{1}{c|}{47.9} & 64.4 & 71.1 & \multicolumn{1}{c|}{67.8} & 56.1 & 56.1 & 56.1 & 18.9  & 21.7 & 20.3 & \cellcolor{white!10}48.0 \\
    \textbf{S5}     & 46.2 & 52.3 & \multicolumn{1}{c|}{49.3} & 52.2 & 59.3 & \multicolumn{1}{c|}{55.8} & 68.2 & 68.2 & 68.2 & NA  & NA & NA & \cellcolor{white!10}57.8 \\
    \hline
    \rowcolor{gray!10} \multicolumn{14}{c}{\textbf{Underlying LLM Alone}} \\
    \textbf{GPT-4o-mini Alone}     & 42.3 & 48.9 & \multicolumn{1}{c|}{45.6} & 66.1 & 74.6 & \multicolumn{1}{c|}{70.4} & 42.6 & 44.5 & 43.5 & 31.1  & 36.7 & 33.9 & \cellcolor{white!10}48.4 \\
    \hline
    
\end{tabular}
    \label{tab:different_LLM_and_role}
\end{table*}

%% file: sections/5_RQ3.tex
\subsection{RQ3: Ablation Study and Impact Analysis}

In this RQ, we examine the contribution of two key components in \TOOL: 1) the Strategy Design, and 2) the Dynamic Team Selection. 
We also analyze the impact of different design choices.

\vspace{0.1cm}
\noindent\textbf{Effectiveness of Dynamic Team Selection.}
The row labeled \textit{``wo Team Selection''} in Table~\ref{tab:RQ1_res} shows the performance of \TOOL when all strategies are combined using simple ensembling. We find that the dynamically selected teams perform slightly better than the full ensemble. This demonstrates that our approach can effectively identify strong strategy combinations while also reducing the costs of using LLMs. 
Overall, the team selection mechanism reduces LLM cost by about 50\% without sacrificing performance. The selected team improves over the full ensemble, possibly because some strategies are less suitable for these tasks, and our dynamic team selection mechanism helps avoid choosing those less effective strategies.

\vspace{0.1cm}
\noindent\textbf{Effectiveness of Strategy Design.}
To demonstrate the effectiveness of our strategy design, we compare two variants of \TOOL. The first variant removes the team selection component, while the second removes both team selection and our custom strategies, which simply prompts the LLM with: ``Please assign a correctness score to the given input data.'' The performance gap between these two variants highlights the contribution of our strategy design.
The last two rows of Table~\ref{tab:RQ1_res} show the results of these two variants. Our strategy design leads to an average improvement of 27.9\% across three tasks, underscoring its critical role in \TOOL's overall effectiveness.

\vspace{0.1cm}
\noindent\textbf{Effectiveness of Individual Strategies.}
Table~\ref{tab:different_LLM_and_role} presents the performance of the five individual strategies (S1–S5) on each dataset. The results show that while individual strategies may perform well or poorly on specific datasets, SE-Jury, as a combined, suitable team, delivers strong and stable performance across all tasks. When comparing average performance across tasks, SE-Jury outperforms S1 (the average of its two variants), S2, S3, S4, and S5 by 14.4\%, 11.4\%, 10.3\%, 34.0\%, and 11.2\%, respectively. This robust performance arises from combining the strengths of the individual strategies, with each strategy contributing to at least one of the best-performing teams across different datasets. Among the individual strategies, S3 performs best overall.

 \vspace{0.1cm}
\noindent\textbf{Comparison between \TOOL and the LLM alone.}
To clearly illustrate the performance of \TOOL relative to its underlying LLM, GPT-4o-mini, we compared them on each dataset. In this comparison, GPT-4o-mini uses the prompt of the Vanilla LLM baseline.
Table~\ref{tab:different_LLM_and_role} shows that SE-Jury consistently and substantially outperforms GPT-4o-mini alone, with a relative improvement of 32.9\% on average. This demonstrates that our proposed design substantially enhances the performance of the LLM.

\begin{table}[b]
\centering
\vspace{-0.3cm}
    \caption{Impact of varying numbers of annotated samples}
   \vspace{-0.1cm}
\resizebox{0.98\linewidth}{!}{%
\begin{tabular}{@{}lrrrrrr@{}}
\toprule
Impact of N Examples & N=2  & N=6  & N=18  & N=20 & N=50 \\ \midrule
Averaged Correlation & 58.1 & 61.1 & 61.1 & 64.3 &  \textbf{64.4} \\ \bottomrule
\end{tabular}
}
\label{varying_number}
\end{table}

\vspace{0.1cm}
\noindent\textbf{Impact of Underlying LLM.}
Impact of Underlying LLM. To assess the impact of the underlying LLM, we also experimented with another state-of-the-art and cost-effective model, DeepSeek-Chat. Table~\ref{tab:different_LLM_and_role} shows that SE-Jury achieves slightly higher effectiveness with DeepSeek-chat (average correlation score: 64.9) compared to GPT-4o-mini (average correlation score: 64.3). These results suggest that SE-Jury can maintain strong performance across different state-of-the-art LLMs.

\vspace{0.1cm}
\noindent\textbf{Impact of Varying Number of Annotated Samples.}
In our team selection module, \TOOL requires a small number of annotated examples (20 samples) from the evaluation dataset to identify the most effective team for each dataset.
To assess the impact of varying the number of annotated examples, we conduct experiments using different values of $N$. For benchmarks with fewer than 200 samples, we set N to at most 20, since a larger N would leave too few remaining test samples and distort the original distribution.
As shown in Table~\ref{varying_number}, while performance may fluctuate, it exhibits a general upward trend as more labeled data is added.
However, asking human annotators to label a large number of samples can be burdensome. 
We believe that the chosen 20 samples mark is a good trade-off between developers’ time to annotate and performance improvement.

\vspace{0.3cm}
\noindent
\begin{tcolorbox} [boxrule=0.8pt,
                top=0.2pt,
                  bottom=0.2pt]
    \textbf{Answer to RQ3}: 
The key components of \TOOL are crucial to its effectiveness. The dynamic team selection mechanism helps identify effective strategy combinations while also reducing LLM usage by around 50\%. It maintains performance and even improves it in some cases. In addition, our strategy design yields an average performance gain of 27.9\%, underscoring its substantial contribution to \TOOL's effectiveness.
\end{tcolorbox}

%% file: sections/5_RQ4.tex
\subsection{RQ4: Correlation to Test Case Execution Outcomes}

In this RQ, we evaluate the correlation between SE-Jury’s scores and test execution outcomes. Additionally, we compare \TOOL with the state-of-the-art LLM judge, CodeJudge, since CodeJudge was primarily evaluated based on its correlation with test execution outcomes in its original paper.

\input{sections/tables/generalizability}

\begin{table}[b]
    \centering
    \vspace{-0.4cm}
    \caption{Comparison with CodeJudge~\cite{tong2024codejudge}.}
    \vspace{-0.2cm}
    \resizebox{0.98\linewidth}{!}{%
    \begin{tabular}{lc c c c c c c c}
        \toprule
         & \multicolumn{2}{c}{\textbf{HumanEval-X}} & \multicolumn{2}{c}{\textbf{CoNaLa}} & \multicolumn{2}{c}{\textbf{APPS}} & \multicolumn{2}{c}{\textbf{BigcodeBench}} \\

         \textbf{Metrics} & $\tau$  & $r_s$  & $\tau$  & $r_s$  & $\tau$  & $r_s$ & $\tau$  & $r_s$ \\ 
        \midrule
        CodeJudge  &\textbf{62.9}   &\underline{62.9}      &54.9    &64.5    &\underline{39.5}    &\underline{39.5}  &\underline{39.2}    &\underline{39.2}   \\
        CodeJudge w/o reference &52.6   &52.6      &\underline{55.3}    &\underline{65.5}    &16.5    &16.5   &31.7    &31.7 \\
        \midrule

        \textbf{\TOOL}    &\textbf{62.9}  &\textbf{68.3}     &\textbf{60.0}    &\textbf{66.9}    &\textbf{44.0}    &\textbf{50.8}  &\textbf{42.0}    &\textbf{46.3}   \\
        \bottomrule
    \end{tabular}
    }
    \label{tab:compare_with_codejudge}
\end{table}

\noindent\textbf{Generalizability to Test Case Execution Outcomes.}
We examine the generalizability of \TOOL to labels based on test execution, where 0 indicates test failure while 1 represents test success.
To this end, we select two popular code generation datasets that include accompanying test cases: HumanEval-X~\cite{HumanEval-X}, which spans multiple programming languages, and APPS~\cite{APPS}, which provides more complex and challenging coding tasks.
Table~\ref{tab:test_based_res} presents the average correlation between \TOOL's evaluation scores and test case execution results on both datasets. 
Table~\ref{tab:test_based_res} demonstrates that \TOOL achieves the highest average correlation with test case execution outcomes, consistently outperforming all baseline methods.
On the HumanEval-X dataset, \TOOL outperforms the best-performing baseline by an average of 14.4\% across five programming languages.
On the APPS dataset, \TOOL achieves even greater improvements, surpassing the best-performing baseline by 71.1\% in terms of average correlation scores. These results confirm \TOOL's generalizability from human-annotated scores to test case-based execution outcomes.

\vspace{0.1cm}
\noindent\textbf{{Comparison of \TOOL with CodeJudge.}}
CodeJudge~\cite{tong2024codejudge} is a state-of-the-art LLM-based evaluator of code correctness in code generation. We compare SE-Jury with CodeJudge using four code generation datasets from CodeJudge’s evaluation suite: HumanEval-X~\cite{HumanEval-X}, APPS~\cite{APPS}, BigCodeBench~\cite{Bigcodebench}, and CoNaLa~\cite{DBLP:conf/msr/YinDCVN08}. Table~\ref{tab:compare_with_codejudge} presents the experimental results. SE-Jury consistently and substantially outperforms CodeJudge, achieving 9.6\% higher average correlation scores across all datasets compared to CodeJudge, and 36.8\% higher than the CodeJudge variant without reference code. In addition, SE-Jury supports a broader range of tasks, including APR and code summarization, which are not supported by CodeJudge.

\vspace{0.3cm}
\noindent
\begin{tcolorbox} [boxrule=0.8pt,
                top=0.2pt,
                  bottom=0.2pt]
    \textbf{Answer to RQ4}: 
SE-Jury achieves the highest alignment with test outcomes, consistently outperforming all baseline methods. Specifically, \TOOL surpasses the state-of-the-art CodeJudge by 9.6\% on average. 
\end{tcolorbox}

%% file: sections/tables/generalizability.tex
\begin{table}[]
    \centering
    \caption{Experimental results for correlation with test case execution outcomes. The highest correlation is highlighted in bold, and the second highest is underlined.}
    \vspace{-0.2cm}
    \resizebox{0.98\linewidth}{!}{%
    \begin{tabular}{l|c c c c c|c}
        \toprule
         & \multicolumn{5}{c}{\textbf{HumanEval-X}} & \multicolumn{1}{|c}{\textbf{APPS}}\\
         
         & \multicolumn{1}{c}{\textit{Python}} & \multicolumn{1}{c}{\textit{Java}} & \multicolumn{1}{c}{\textit{C++}} & \multicolumn{1}{c}{\textit{JavaScript}} & \multicolumn{1}{c}{\textit{Go}} & \multicolumn{1}{|c}{}\\

         \textbf{Metrics} & {$avg\text{-}cor$}  & {$avg\text{-}cor$}  & {$avg\text{-}cor$}  & {$avg\text{-}cor$}  & {$avg\text{-}cor$}  & {$avg\text{-}cor$}  \\ 
        \midrule

        BLEU        &48.0   &30.6      &47.7    &39.0    &34.5    &1.1     \\
        ROUGE-L     &47.6   &33.2      &45.7    &42.8   &33.4    &2.0     \\
        METEOR      &51.9   &37.4      &46.2    &45.4   &32.1    &10.7     \\
        ChrF++      &47.3   &35.9      &47.5    &44.4   &30.8    &1.9    \\
        CodeBLEU    &52.5   &41.4     &43.7    &47.5    &31.8   &20.5     \\
        RUBY        &42.1   &34.8     &44.9    &42.2    &32.5    &4.0     \\
        CrystalBLEU &36.7   &30.0     &42.4    &39.7  &31.3    &3.2     \\
        MoverScore  &42.1   &32.9      &42.1    &38.8   &42.3    &1.8     \\
        BERTScore  &43.1   &34.2     &45.3    &41.7    &26.8    &3.9     \\
        CodeBERTScore &46.1   &37.7      &48.1    &42.1  &37.2    &1.7    \\
        Vanilla LLM  &\underline{62.9}   &\underline{59.1}      &57.4    &\underline{62.2}    &45.7    &16.0   \\
    
        ICE-Score &60.1   &53.0      &\underline{59.3}    &60.5    &\underline{53.9}    &\underline{27.7}    \\
        \midrule
        \textbf{\TOOL}    &\textbf{74.0}  &\textbf{63.6}     &\textbf{72.2}    &\textbf{62.3}    &\textbf{55.9}    &\textbf{47.4}     \\
        \bottomrule
    \end{tabular}
    }
    \vspace{-0.5cm}
    \label{tab:test_based_res}
\end{table}

%% file: sections/6_discussion.tex
\vspace{-0.2cm}
\section{Discussion\label{discussion}}

\vspace{-0.1cm}
\subsection{Limitations}
First, although \TOOL generally produces evaluation results that align well with human judgments, it still encounters some failure cases, especially in the code summarization task. The evaluation of code comments is often subjective, and even human annotators may disagree on what constitutes a correct summary. Similarly, when dealing with ambiguous or controversial comments, \TOOL may find it difficult to generate scores that match human assessments. Nevertheless, this limitation is common across all baseline methods, and \TOOL still demonstrates better performance compared to other approaches.
Second, \TOOL incorporates multiple strategies (prompts) that reflect different perspectives and it relies on commercial LLMs, which results in an additional cost of by around US\$ 0.1 per 100 samples. To manage this, we use OpenAI’s GPT-4o Mini~\cite{openai_gpt4o_mini} as the backbone model, which is one of the most cost-effective commercial LLMs. We also introduce a team selection mechanism to further reduce the number of LLM queries. Although traditional automatic metrics such as BLEU are free, they often fail to accurately reflect human judgment~\cite{DBLP:journals/jss/EvtikhievBSB23} and can lead to misleading conclusions.

\vspace{-0.1cm}
\subsection{Threats to Validity}
Our findings are limited to the specific SE datasets examined in this study and may not generalize to all SE datasets. To mitigate this limitation, we selected three widely adopted datasets—CoNaLa, Card2Code, and Summary-Assess—and introduced APR-Assess, a human-annotated dataset not previously studied in this context. These benchmarks collectively span three major SE tasks: code generation, automated program repair, and code summarization. They also cover diverse software artifacts, including source code, code changes, and code comments.
Another potential threat to validity is our choice of evaluation metrics. Following prior work~\cite{DBLP:journals/jss/EvtikhievBSB23,ICE-Score}, we adopt widely used metrics for this task: Kendall’s $\tau$ and Spearman’s $r_s$ for measuring statistical correlation. We employ the widely used Cohen’s Kappa score to quantify inter-rater agreement.
Moreover, \TOOL uses the OpenAI GPT-4o mini model for its cost-effectiveness, though its performance may improve when more advanced LLMs (e.g., GPT-4.5) are used. 

%% file: sections/7_related_work.tex
\vspace{-0.3cm}
\section{Related Work\label{related}}

\noindent\textbf{Traditional Metrics for Correctness Evaluation.}
Many SE-specific metrics have been proposed to assess the correctness of SE data. 
Ren et al. introduced CodeBLEU~\cite{DBLP:journals/corr/abs-2009-10297}, which extends BLEU by incorporating code syntax and structure. Tran et al. proposed RUBY~\cite{DBLP:conf/iwpc/TranTNNN19}, which measures similarity using lexical, syntactic, and semantic representations. Eghbali et al. developed CrystalBLEU~\cite{DBLP:conf/kbse/EghbaliP22}, which improves BLEU by filtering out common n-grams to focus on more informative patterns.
Zhou et al. proposed CodeBERTScore~\cite{DBLP:conf/emnlp/Zhou0AN23}, which adapts BERTScore to code by leveraging a fine-tuned CodeBERT model. 
Mastropaolo et al. introduced SIDE~\cite{SIDE}, a metric that applies contrastive learning to enhance cosine similarity-based evaluation on code summaries.
Additionally, Evtikhiev et al.~\cite{DBLP:journals/jss/EvtikhievBSB23} empirically evaluated six metrics, such as BLEU and ROUGE-L, against human judgments and found significant misalignment, emphasizing the need for more accurate and reliable automatic metrics.
In contrast to prior traditional metrics, SE-Jury leverages LLMs’ reasoning and code comprehension abilities for semantic evaluation. 

\vspace{0.1cm}
\noindent\textbf{LLM-based Metrics for SE Artifacts Correctness Evaluation.} 
Zhuo et al. proposed ICE-Score\cite{ICE-Score}, which prompts an LLM to assign a correctness score based on predefined criteria. 
CodeScore~\cite{dong2025codescore} assesses the functional correctness of generated code by estimating its pass ratio (the proportion of unit tests the code passes) and executability (whether the code can be executed successfully).
Yang et al.~\cite{yang2025codeditingreasoningbasedmetricfunctional} introduced Code-DiTing, a judge model enhanced by reasoning distillation from DeepSeek-R1, designed to align closely with test case execution outcomes.
CodeJudge~\cite{tong2024codejudge} is the state-of-the-art LLM judge, which uses the LLM to evaluate the functional correctness of generated code in code generation.  
The above methods rely on a single prompting strategy and are limited to evaluating code correctness.
In contrast, SE-Jury introduces multiple strategies and supports a broader set of tasks, including code generation, summarization, and automated program repair. We also introduce a dynamic team selection module that enhances performance while reducing LLM API costs. Moreover, SE-Jury substantially outperforms both CodeJudge and ICE-Score. Furthermore, we demonstrate that SE-Jury can achieve agreement levels with human annotators that are close to the inter-annotator agreement observed among humans on code generation and automated program repair tasks.
Moreover, Kumar et al.~\cite{kumar2024llms} introduced an LLM judge for the bug summarization task, relying on a single evaluation strategy. In contrast, our work addresses multiple SE tasks and introduces five distinct prompt strategies along with a dynamic team selection module to enhance performance and reduce LLM API costs.
Lastly, Chun et al.~\cite{chun2025multi} introduced a multi-agent system designed to generate high-quality code summarization and translation, whereas our work focuses on assessing the correctness of AI-generated SE artifacts.

\vspace{0.1cm}
\noindent\textbf{LLM-based Methods for Evaluating Non-functional Properties of SE Artifacts.}
Wang et al.~\cite{wang2025can} empirically investigate LLM-as-a-judge methods from NLP for evaluating SE tasks, focusing on consistency and readability aspects.
CodeUltraFeedback~\cite{weyssow2024codeultrafeedback} is a recent study focused on non-functional properties, specifically coding preferences such as style or readability. The study first collects a preference dataset consisting of instructions and corresponding responses generated by various LLMs. It then introduces an LLM-based judge to assess code preferences using this dataset, achieving competitive performance.
In contrast, SE-Jury primarily evaluates functional correctness.
Therefore, we consider our work orthogonal to CodeUltraFeedback~\cite{weyssow2024codeultrafeedback} and Wang et al.~\cite{wang2025can} and view extending SE-Jury to non-functional properties as an interesting direction for future work.

\vspace{0.1cm}
\noindent\textbf{Fine-tuned LLM Judges from Natural Language Processing.}
Several fine-tuned LLM judges have been proposed in Natural Language Processing (e.g., \cite{ligenerative}, \cite{kim2024prometheus}). These approaches require large-scale training datasets, but annotated datasets for evaluating the correctness of AI-generated SE artifacts are still scarce. This scarcity makes it difficult to apply fine-tuned judges in the SE domain. Furthermore, to the best of our knowledge, no SE-specific fine-tuned LLM judge has been published. 
Given these limitations inherent to the field, we do not compare SE-Jury with fine-tuned evaluators in our experiments. 
In contrast, SE-Jury performs effectively with only 20 annotated samples, making it a more practical solution.

%% file: sections/8_conclusion.tex
\vspace{-0.1cm}
\section{Conclusion and Future Work\label{conclusion}}
In this paper, we present \TOOL, the first LLM-as-Ensemble-Judge metric specifically designed to evaluate the correctness of generated software artifacts.
\TOOL defines five distinct evaluation strategies and introduces a team selection mechanism that chooses an appropriate subset of strategies.
Our experiments show that \TOOL achieves substantially and consistently higher correlations with human judgments, with improvements ranging from 29.6\% to 140.8\% on average compared to existing automatic evaluation metrics. Furthermore, \TOOL reaches agreement levels with human annotators that are close to inter-annotator agreement in tasks like code generation and automated program repair. 

In future work, we plan to extend \TOOL beyond correctness evaluation to cover additional dimensions, such as assessing non-functional properties of generated software.

\vspace{0.2cm}
\noindent \textbf{Acknowledgement.}  This research was supported by the Singapore Ministry of Education (MOE) Academic Research Fund (AcRF) Tier 1 grant (Project ID: 23-SOL-SMU-004). Any opinions, findings and conclusions or recommendations expressed in this material are those of the author(s) and do not reflect the views of the Ministry of Education, Singapore.